\documentclass{aastex631}

\usepackage{booktabs}
\begin{document}

\title{Free-Free Gaunt factors for atmospheres of accreting pulsars observable with X-ray space missions}

\correspondingauthor{Parisee S. Shirke}
\email{parisee@iucaa.in}

\author[0000-0003-2977-3042]{Parisee S. Shirke}
\affiliation{Inter-University Centre for Astronomy and Astrophysics \\
Post Bag 4, Ganeshkhind, \\
Pune 411 007, India}

\begin{abstract}

Free-Free Gaunt factors for X-ray absorption by hot plasma in the presence of a strong magnetic field are reported. Modified formulae are used for application to the non-local thermodynamic equilibrium conditions found in accreting pulsar atmospheres. Given upcoming global X-ray polarimetric space missions, these can be used for the construction of an absorption matrix in discrete-ordinate polarised radiative transfer.
\end{abstract}


\section{Introduction} \label{sec:Intro}

The radiation scattering environment in a pulsar atmosphere is conventionally taken to be an accreted hot, tenuous, homogeneous plasma comprising mostly ionized hydrogen with trace helium \citep{MeszarosNagel1985a,MeszarosNagel1985b}. Gaunt factors may need to be computed to account for photon-plasma interactions in the presence of a strong magnetic field using appropriate absorption and scattering cross-sections \citep{hardingdaugherty1991} for determining the effective opacity and optical depth (See e.g. \cite{Meszaros1988} and references therein). The plasma consists of three different values of co-existent temperatures, namely (i) the plasma ion temperature, (ii) the electron temperature parallel to the magnetic field $\vec{B}$ and (iii) the temperature representing the electron distribution over the Landau levels, which results in non-thermal local equilibrium (non-LTE) conditions. Continuous absorption is dominated by the free-free Brem\ss trahlung process except near the cyclotron resonance (where resonant magneto-Compton scattering may set in). 

Gaunt factors are slowly varying smooth functions that appear as multiplicative quantum mechanical corrections in the description of continuous (free-free or bound-free) absorption and emission \citep{1979rpa..book.....R} and were developed from the classical formula by \cite{kramers1923xciii}. These were introduced to provide a better description of stellar opacity arising from highly ionized plasma interactions in stellar interiors \citep{gaunt1930continuous} and were later named after their discoverer by \cite{1939isss.book.....C}. \cite{1961ApJS....6..167K} have summarized the analytic formulae for computation of Gaunt factors at non-relativistic energies. Although recent works provide improved calculations of free-free Gaunt factors for astrophysical applications like radiative transfer in galaxy clusters, spectral distortions of CMB, and free-free galactic foreground \citep{2020MNRAS.492..177C}, this work computes the same in the context of accretion-powered pulsars.

\section{Methods}\label{sec:methods}

\begin{figure}
    \centering 
    \includegraphics[clip, trim = 0 35 0 35, width=.8\textwidth]{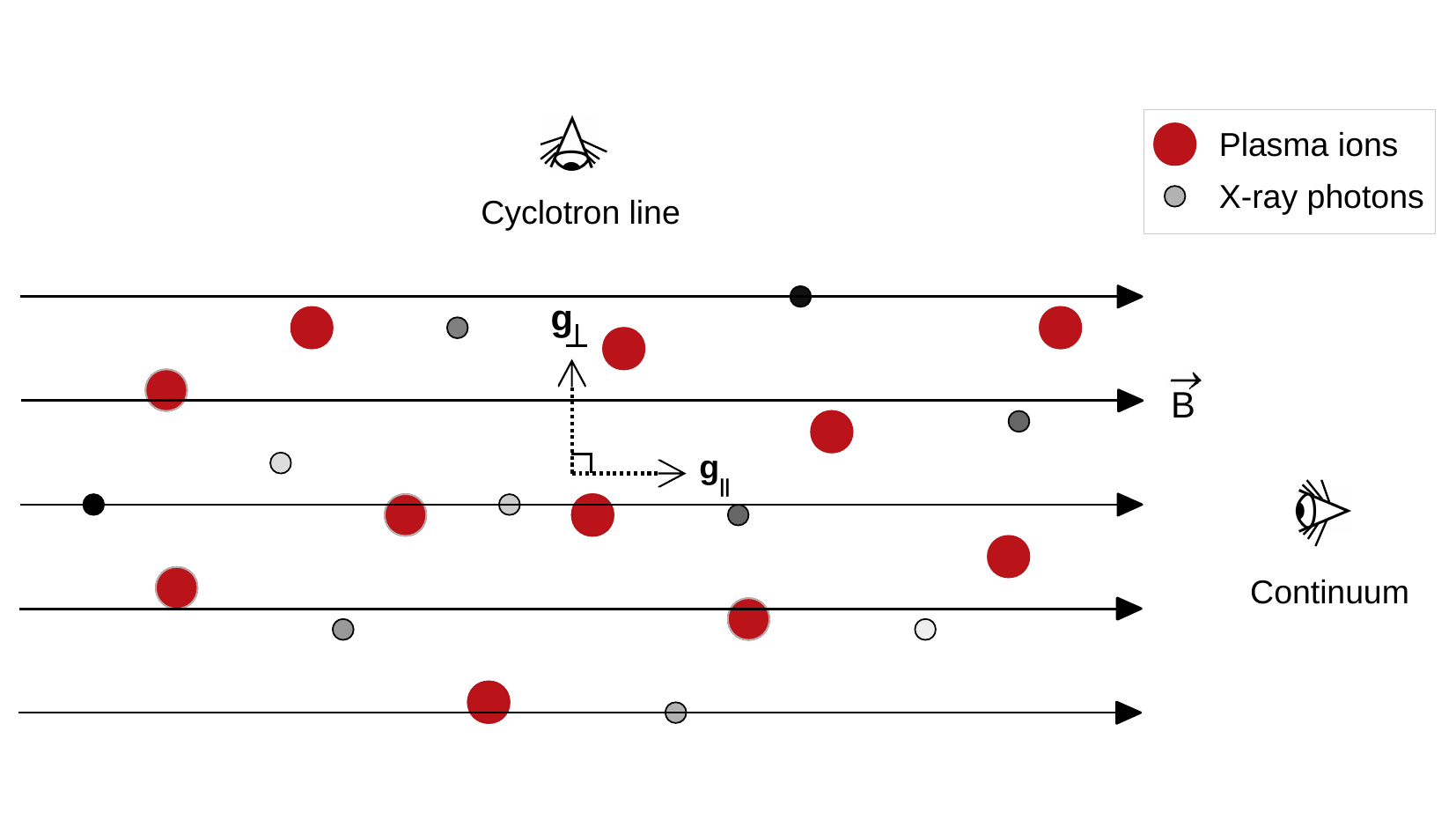}
    \caption{Sketch of a model pulsar atmosphere with a combination of accreted plasma ions and emitted thermal Brem\ss trahlung X-ray photons. The direction of the dipolar magnetic field is perpendicular to the surface near the polar regions. Two gaunt factors $(g_{\perp}$ and $g_{||})$ are required to characterize the anisotropic X-ray photon absorption in the presence of a strong magnetic field. These correspond to the cyclotron line and X-ray continuum, respectively.}
    \label{fig:diagram_plasma}
\end{figure}
Gaunt factors appear in the expression for the magnetic free-free absorption coefficient (See \cite{MeszarosNagel1985a, MeszarosNagel1985b, Ventura1979}),
    \begin{equation} 
        \alpha_i(\omega, \theta) = \alpha_0 [ \phi_+(\omega)|e^i_+|^2 g_{\perp} + \phi_-(\omega)|e^i_-|^2 g_{\perp} +\phi_z(\omega)|e^i_z|^2 g_{||}] 
    \end{equation}
    where $\alpha_0$ is the non-magnetic free-free absorption coefficient measuring the free-free opacity of non-magnetized plasma given as,
    \begin{equation} \label{alpha_0}
        \alpha_0=\frac{4\pi^2Z^2\alpha^3\hbar^2c^2N}{m_e \omega^3 \sqrt{\pi k T / 2m_e}},
    \end{equation}
    and 
    \begin{equation}
        \phi_{+,-,z}=\frac{3}{2} \frac{c}{r_0 \omega} | \text{Im } T_{+,-,z}| ,
    \end{equation}
where $Z$ represents the mean plasma atomic number, $\alpha$ is the fine structure constant $(=1/137)$, $T$ is the plasma temperature ($=8$ keV, $10$ keV for plasma in model pulsar atmospheres) \citep{MeszarosNagel1985b}, $N$ is the number of electrons and $\omega$ is the incident spectral energy, $T_{+,-,z}$ are the components of the polarization tensor, $r_0$ is the electron classical radius along with standard physical constants.

To account for anisotropic effects due to the magnetic field, the Gaunt factors are resolved into two components as shown in Fig. \ref{fig:diagram_plasma}: (i) for the cyclotron line  $g_{\perp}$ (See \citet{Nishimura2008} for details on cyclotron line profiles) and (ii) the continuum $g_{||}$. The direction of the magnetic field is the natural axis with respect to which the directions in the suffixes are defined. The modified expressions to incorporate a non-LTE situation \citep{1980ApJ...236..904N} are,
 \begin{equation}
        g_\perp (\omega, \omega_c, T_{||}) = \int\limits_{-\infty}^{\infty} C_1 \Bigg( \frac{\omega}{\omega_c} e^{2x}  \Bigg) e^{-\frac{\hbar \omega}{k T_{||}} sinh^2x}  dx
    \end{equation}
    \begin{equation}
        g_{||} (\omega, \omega_c, T_{||}) = 2 \int\limits_{-\infty}^{\infty} \Bigg( \frac{\omega}{\omega_c} e^{2x} \Bigg) C_0 \Bigg( \frac{\omega}{\omega_c} e^{2x}  \Bigg) e^{-\frac{\hbar \omega}{k T_{||}} sinh^2x} dx
    \end{equation}
    where $C_1$ and $C_0$ are Coulomb matrix elements defined by \cite{1975NCimB..26..537V}, representing the strength of transition from the $n^{\text{th}}$ to $n'^{\text{th}}$ Landau level. These values are tabulated in \cite{1973PhysRevA.8.3021V} as $W_{00}$ and $W_{10}$ for parameter values ranging from $z= 0.01 - 10.00$, with $z$ being defined as,
    \begin{equation}
       z= \frac{1}{4 \gamma} [\beta^2 + (\Delta K)^2]
    \end{equation}
    where, $\gamma = \frac{1}{2}eB/\hbar c$, $\beta^2/4\gamma$ is the screening parameter and $\Delta K = k'-k$ represents the momentum transfer. 

For a known value of model plasma temperature $T_{||}$, the Gaunt factors are functions of the X-ray spectral energy $\omega$ and the systemic cyclotron energy $\omega_c$, where the latter is fixed for a particular system but can vary from pulsar to pulsar. Using $kT=8$ keV for model accreting pulsar atmospheres (See Appendix), the integrals are computed\footnote{Using dimensional analysis, $\hbar$ is hereby omitted from the exponent in the implementation. Both the numerator, $\omega$ and denominator, $kT$ are in units of keV. The cyclotron energy, $\omega_c$ in the following expression for $z$ is also in keV.} using $z = \frac{\omega}{\omega_c} e^{2x}$ and reading off the corresponding value of $W_{00}$ and $W_{10}$ for each $x$, using linear interpolation, if required. The sufficiency of the range of \citeauthor{1973PhysRevA.8.3021V}'s $z$ values was separately confirmed by verifying that the integrands have vanishingly small values outside this range (by evaluating the Gaunt integrals with Coulomb matrix elements set to fixed constants). 

The cyclotron energies $\omega_c$ in Tables \ref{table:g_perp} and \ref{table:g_para} are initially have discrete sampling across the full range of pulsar magnetic field strengths ($B = 10^{11}-10^{13}$ Gauss). The resultant set of values subsumes the model cyclotron energies of 38 keV and 50 keV used by \cite{MeszarosNagel1985a} and \cite{MeszarosNagel1985b}. On the other hand, the range of spectral energies $\omega$ in Tables \ref{table:g_perp}, \ref{table:g_para} is initially kept consistent with the discrete set of values used in \cite{MeszarosNagel1985a}'s radiative transfer computations.  % for pre-computation towards implementation of the Feautrier method. 
Both are later allowed to continuously span the full observing band of broad-band X-ray space missions, $\omega = 1-100$ keV and pulsar magnetic field strengths $\omega_c =1-120$ keV at a fine spectral resolution of $\Delta \omega = 0.1$ keV. Recursive adaptive Simpson's quadrature is used to perform the integration since it provides results -- with error bars -- consistent with those obtained using other integrators\footnote{\url{https://scipy.org} \citep{virtanen}. See Sec. \textit{Software}.} while using a number of iterations, $N$ close to the default value recommended for the integration module. The preliminary criteria for the choice of $N$ is that a change to $2N$ should not produce a change in the results of Gaunt factor values within $1\%$ accuracy.

\section{Results and Discussions} \label{sec:rnd}

\begin{figure}
    \centering
    \includegraphics[width=.48\textwidth, trim=300 20 270 20, clip]{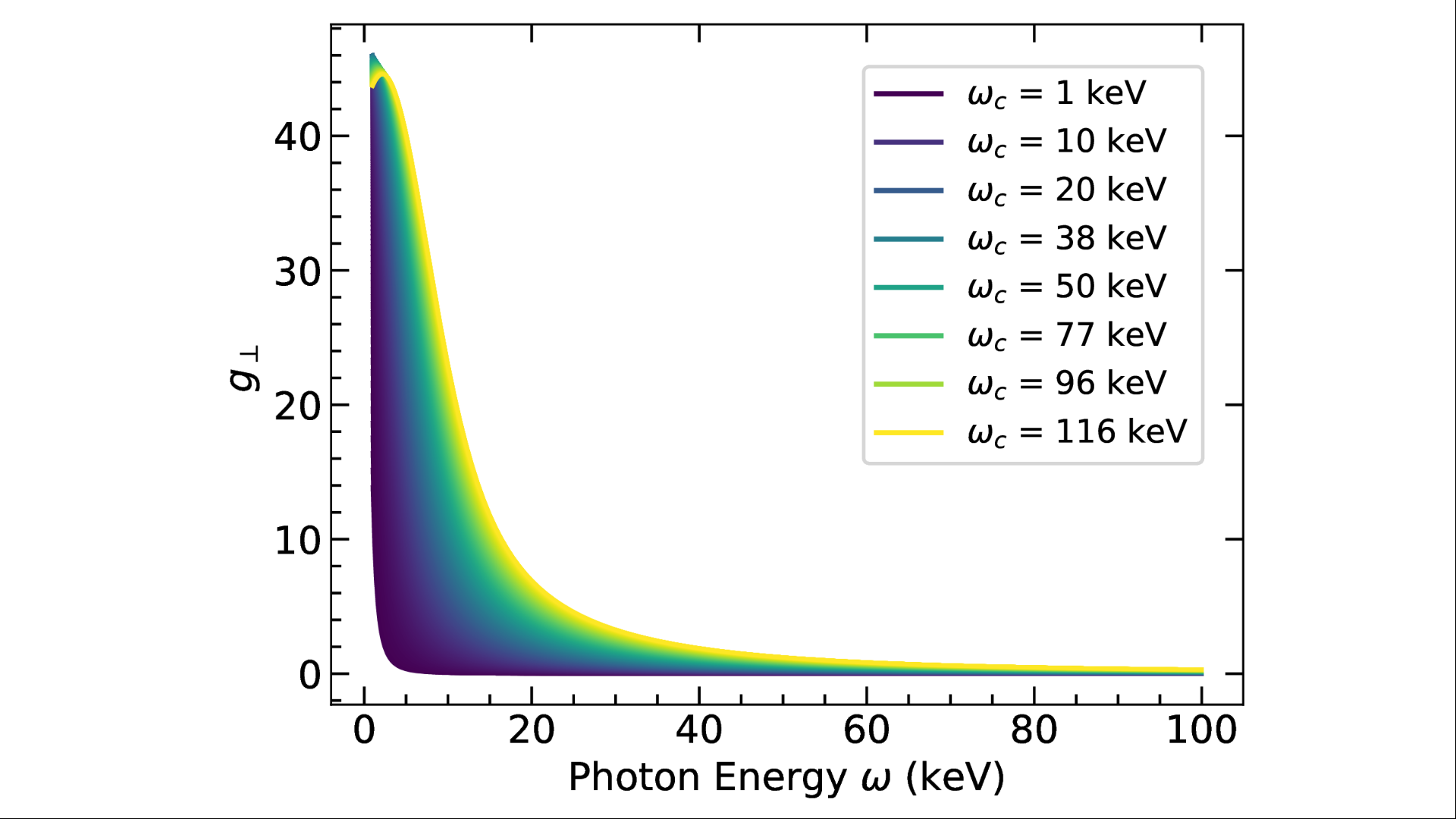} \hfill
    \includegraphics[width=.48\textwidth, trim=300 20 270 20, clip]{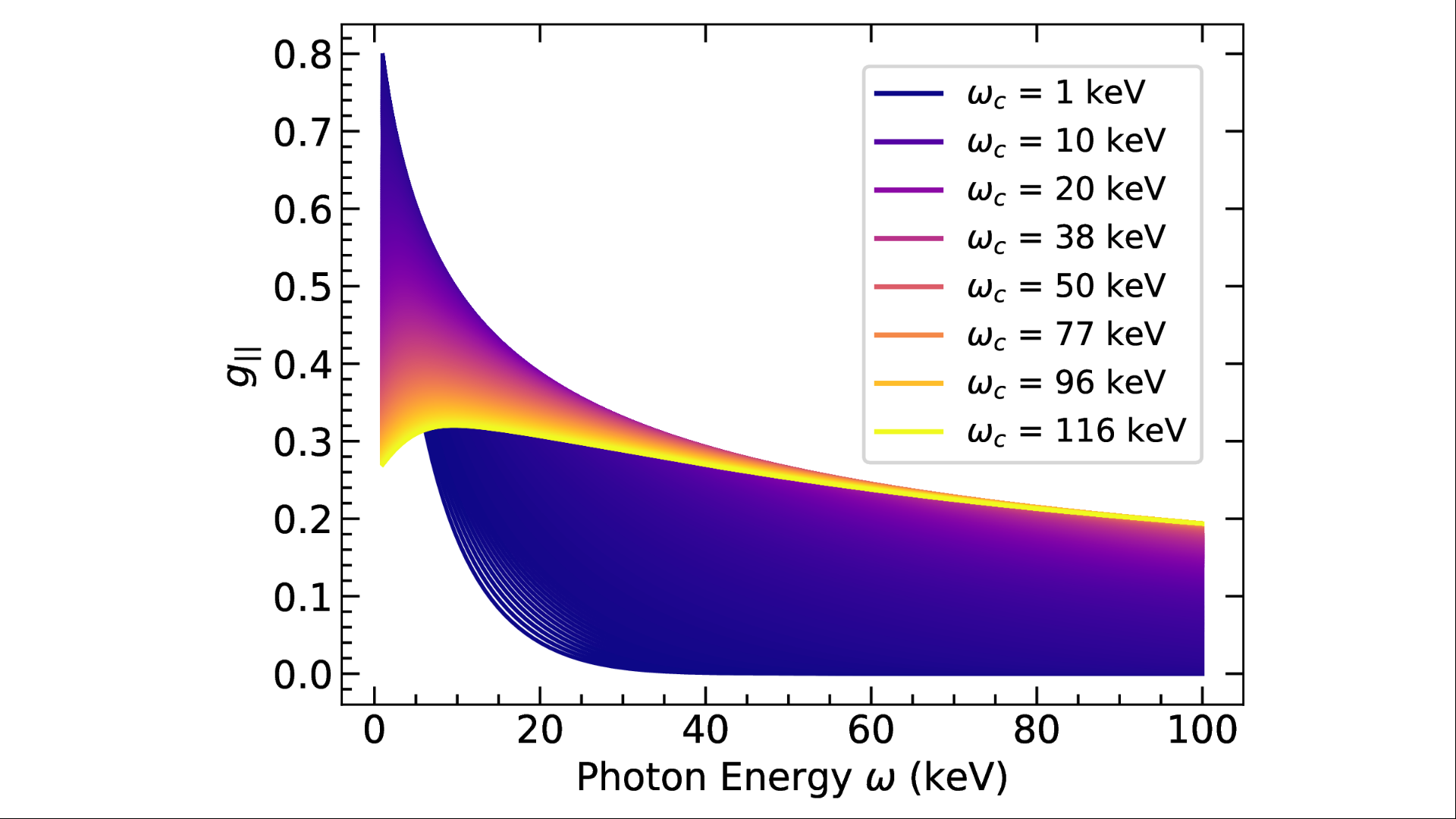}
    \caption{Modified anisotropic Gaunt functions (\textit{left}) $g_{\perp}$ and (\textit{right}) $g_{||}$ for the full spectral range of broad-band X-ray space missions i.e. the full observing range of X-ray photons energies, $\omega = 1-100$ keV  and (\textit{isochors}) the full span of source pulsar magnetic field strengths, $B = 10^{11}-10^{13}$ G sampled at a fine spectral resolution of $\Delta \omega = \Delta \omega_c =0.1$ keV for a plasma temperature of $kT=8$ keV.}
    \label{fig:vary_w}
\end{figure}

Tables \ref{table:g_perp} and \ref{table:g_para} and Fig. \ref{fig:vary_w} show the computation results for the values of Gaunt factors, $g_{\perp}$ for pulsar cyclotron lines and $g_{||}$ for X-ray continua. Since cyclotron resonance scattering features tend to be sharp, the energy band was sampled with a fine resolution ($\Delta \omega =0.1$ keV) for close sampling, especially near the cyclotron line energy ($\omega_c$). As seen in Fig. \ref{fig:vary_w}, the Gaunt factors exhibit their typical smooth behaviour. There are no discontinuities, even near the cyclotron resonant energies which otherwise produce sharp scattering features in X-ray pulsar spectra. Both the Gaunt factors are seen to decrease with the X-ray photon energy as mentioned in \cite{VenturaNagelMeszaros1979}. $g_{\perp}$ increases with an increase in magnetic field strength whereas $g_{||}$ is seen to exhibit the reverse trend, getting suppressed with an increase in the magnetic field strength. It would be a trivial exercise to represent these results -- plotted as per convention in Fig. \ref{fig:vary_w} for maintaining consistency with existing reports in the literature -- as Gaunt functions on a 3-D grid with spectral energy $\omega$ and the cyclotron energy $\omega_c$ as the choice of $X$ and $Y$ axes.
\begin{longtable}{cccccccc}
    \caption{Modified anisotropic Gaunt factor, $g_{\perp}$ for a model \citeauthor{MeszarosNagel1985b} plasma temperature of $8$ keV \textit{(columns)} over the full range of source pulsar magnetic field strengths, $B =10^{11}-10^{13}$ G with discrete sampling for $\omega_c$ including model \citeauthor{MeszarosNagel1985b} values and \textit{(rows)} model \citeauthor{MeszarosNagel1985b} energies for $\omega$.}
    \label{table:g_perp}\\
    \hline \hline
    Energy & \multicolumn{7}{c}{ Cyclotron energy $\lq \omega_c$' } \\
     $\lq \omega$'         & \multicolumn{7}{c}{(keV)}\\ \cmidrule{2-8}(keV) & 1.0 & 20.0 & 38.0 & 50.0 & 77.0 & 96.0 & 116.0 \\ \hline
1.58 & 5.300323 & 41.53 & 44.68945 & 45.2593 & 45.31703 & 44.96964 & 44.471494 \\
3.85 & 0.640098 & 20.74975 & 31.43405 & 35.39241 & 40.17336 & 41.83082 & 42.813269 \\
8.98 & 0.066791 & 4.15618 & 8.714412 & 11.805245 & 18.307305 & 22.234345 & 25.737028 \\
18.37 & 0.006339 & 0.910697 & 2.097411 & 2.940173 & 4.896875 & 6.305912 & 7.811599 \\
29.13 & 0.000714 & 0.33181 & 0.797979 & 1.151917 & 2.015293 & 2.649239 & 3.326339 \\
38.59 & 0.000087 & 0.180185 & 0.440854 & 0.639782 & 1.143135 & 1.528293 & 1.948742 \\
51.71 & 0.000002 & 0.093834 & 0.240337 & 0.349771 & 0.627453 & 0.845317 & 1.09022 \\
84.66 & 0.0 & 0.028196 & 0.084971 & 0.128962 & 0.233738 & 0.314034 & 0.405335 \\
\hline
\end{longtable}
\newpage
\begin{longtable}{cccccccc}
    \caption{Modified anisotropic Gaunt factor, $g_{||}$ for a model \citeauthor{MeszarosNagel1985b} plasma temperature of $8$ keV \textit{(columns)} over the full range of source pulsar magnetic field strengths, $B = 10^{11}-10^{13}$ G with discrete sampling for $\omega_c$ including model \citeauthor{MeszarosNagel1985b} values and \textit{(rows)} model \citeauthor{MeszarosNagel1985b} energies for $\omega$.}
    \label{table:g_para}\\
    \hline \hline
    Energy & \multicolumn{7}{c}{ Cyclotron energy $\lq \omega_c$'} \\
    $\lq \omega$'         & \multicolumn{7}{c}{(keV)}\\ \cmidrule{2-8}
(keV) & 1.0 & 20.0 & 38.0 & 50.0 & 77.0 & 96.0 & 116.0 \\ \hline
1.58 & 0.693441 & 0.580327 & 0.470625 & 0.422493 & 0.3486 & 0.31256 & 0.282872 \\
3.85 & 0.457043 & 0.570444 & 0.483836 & 0.441327 & 0.371953 & 0.33657 & 0.306753 \\
8.98 & 0.202961 & 0.50167 & 0.45561 & 0.427458 & 0.376749 & 0.34888 & 0.324334 \\
18.37 & 0.051592 & 0.398091 & 0.3917 & 0.37875 & 0.348874 & 0.330137 & 0.312663 \\
29.13 & 0.008675 & 0.317857 & 0.33395 & 0.330488 & 0.315196 & 0.303548 & 0.291772 \\
38.59 & 0.001224 & 0.266875 & 0.295343 & 0.296937 & 0.289336 & 0.281716 & 0.273456 \\
51.71 & 0.000038 & 0.213812 & 0.25377 & 0.260534 & 0.259949 & 0.25582 & 0.250702 \\
84.66 & 0.0 & 0.134059 & 0.180778 & 0.196641 & 0.208867 & 0.209444 & 0.208118 \\
\hline 

\end{longtable}

\section{Conclusions} \label{sec:conclusions}
Brem\ss trahlung Gaunt factors  -- used in the studies of electrodynamics and radiative processes -- are presented over the complete broad-band X-ray observing range of $\omega = 1-100$ keV for the full span of accreting pulsar magnetic field strengths, $B = 10^{11}-10^{13}$ Gauss ($\omega_c = 1-120$ keV) with (i) discrete sampling for model values and (ii) continuous sampling at a fine spectral resolution of 0.1 keV for a model plasma temperature of 8 keV. The results are shown in Fig. \ref{fig:vary_w}. 

Since accreting X-ray pulsars are promising non-terrestrial physics laboratories, this, in general, characterizes the physics of magnetically confined (X-ray pulsar) plasmas, particularly, their interactions with incident high energy radiation in vacuum. Exploiting the naturally-occurring condition of a high-velocity accreted plasma funneled by a strong dipolar pulsar magnetic field and interacting with thermal Brem\ss trahlung X-ray radiation, the free-free Gaunt factors for a hot ($T_{||} = 8$ keV), magnetized ($B= 10^{11} - 10^{13}$ G) plasma in non-LTE conditions, including anisotropic effects ($g_{\perp}], g_{||}$) are presented for $1-100$ keV X-ray photon ($\omega$) absorption in model pulsar atmospheres. A new, custom numerical tool \lq \texttt{Gaunt Factor Calculator}' is developed for the same.

\begin{acknowledgments}
The author would like to thank Prof. D. Bhattacharya and Prof. D. Mukherjee for useful discussions, suggestions and ideas and Prof. G. C. Dewangan for support. %\textcolor{gray}{Constructive comments and suggestions from an anonymous referee have improved the content of the paper.}
\end{acknowledgments}

\facilities{Dell OptiPlex 5060 Desktop\footnote{\url{https://www.dell.com/learn/in/en/inbsd1/shared-content~data-sheets~en/documents~optiplex_5060_spec_sheet.pdf}} with an Intel\textsuperscript{\textregistered} Core™ i7-8700T processor, Ubuntu\textsuperscript{\textregistered} 18.04.3 LTS (64-bit) Operating System, 8 GB DDR4 RAM \& 2 TB HDD.}

\software{NumPy\footnote{\url{https://numpy.org}} \texttt{Ver 1.16.4} \citep{harris}, SciPy\footnote{\url{https://scipy.org}} \texttt{Ver 1.3.0} \citep{virtanen}, Matplotlib\footnote{\url{https://matplotlib.org}} \texttt{Ver 3.1.0} \citep{hunter} packages in Jupyter environment \citep{jupyter} for Python \texttt{3} \citep{rossum},
the SAO/NASA Astrophysics Data System\footnote{\url{https://ui.adsabs.harvard.edu/}},
e-Print arXiv\footnote{\url{https://arxiv.org/}}.}

\appendix
\section{Visual representation}
\begin{figure}
    \centering
    \includegraphics[width=0.45\linewidth, clip, trim = 0 0 0 50]{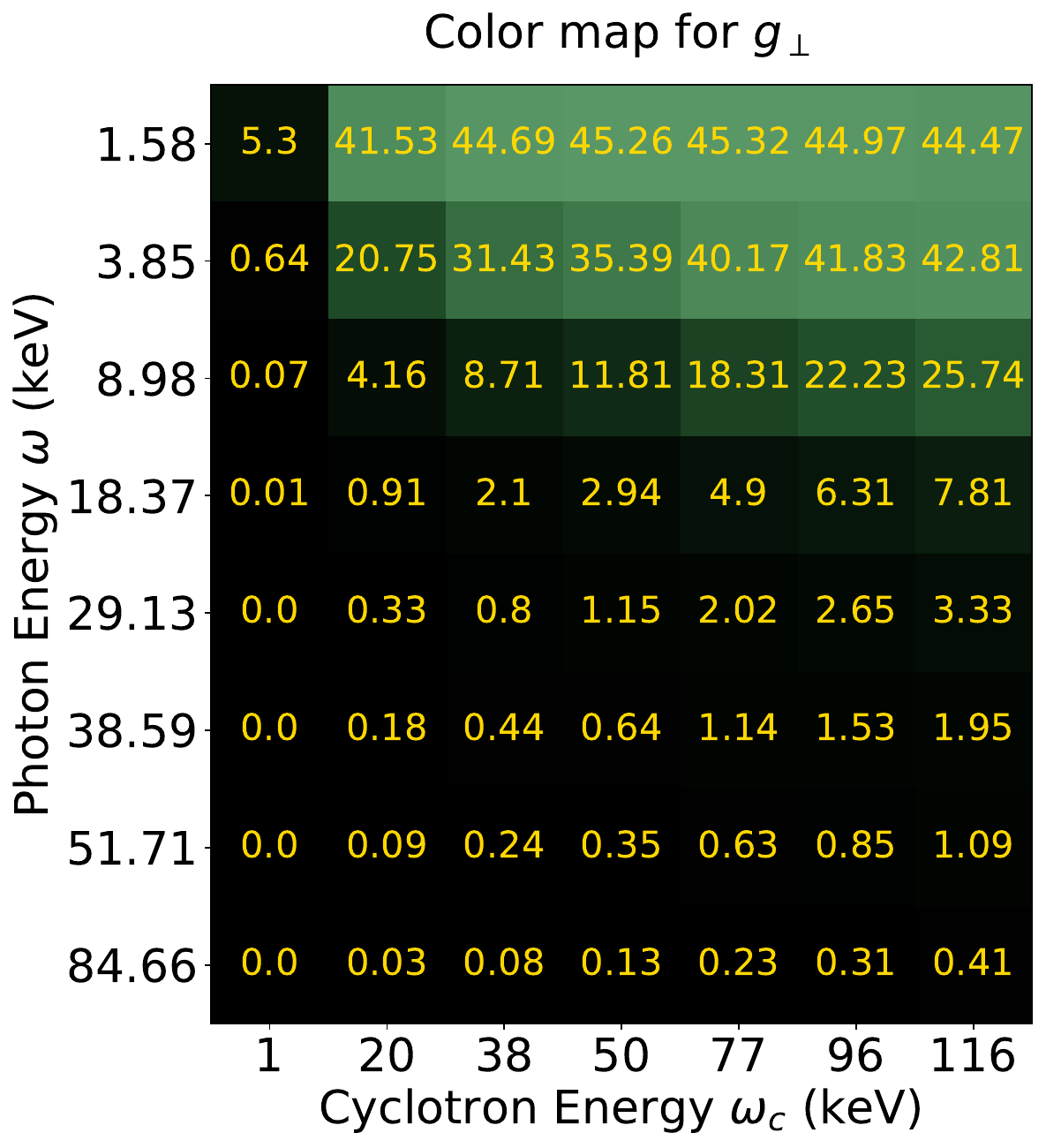} 
    \includegraphics[width=0.45\linewidth, clip, trim = 0 0 0 50]{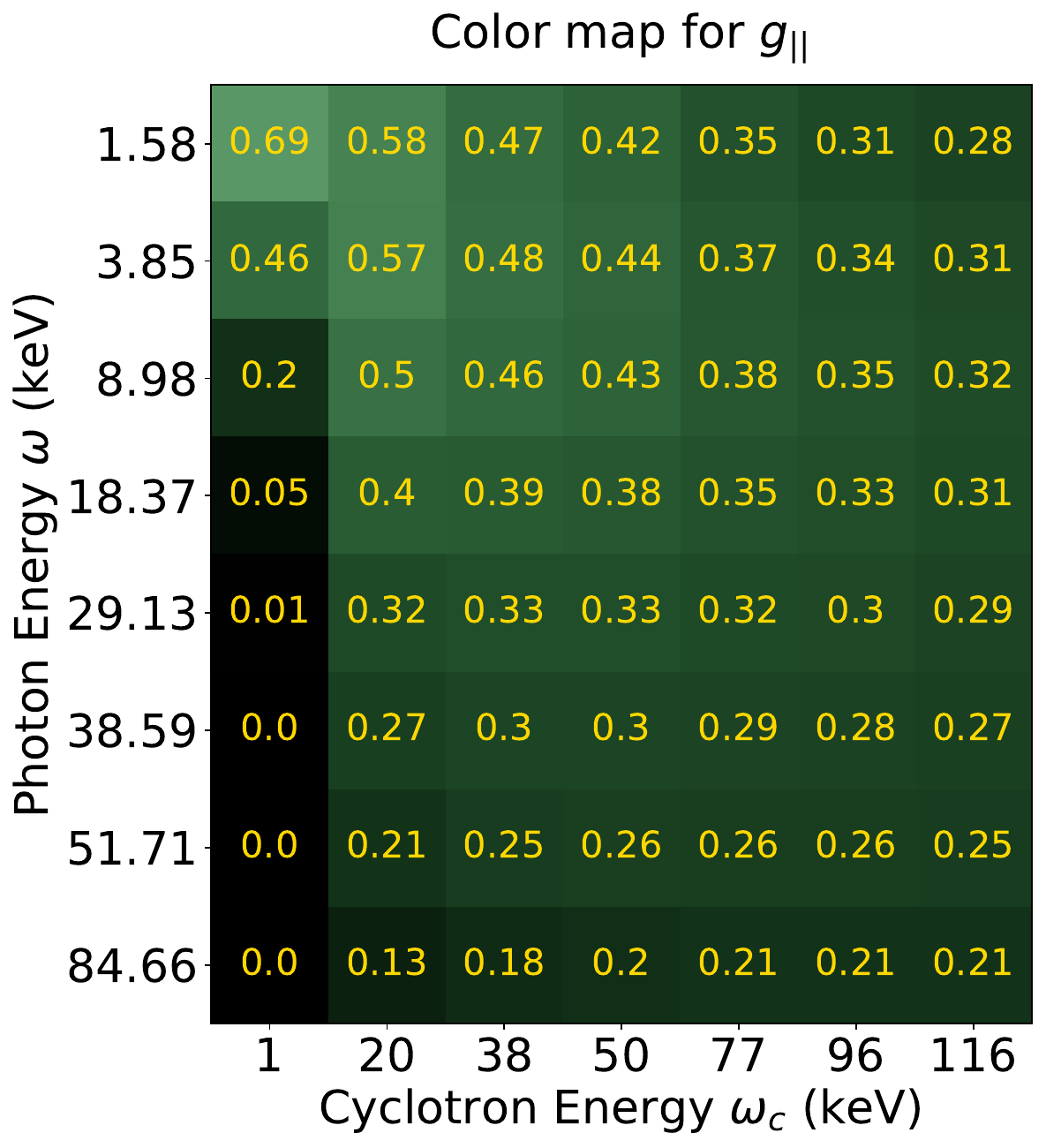}
    \caption{Modified anisotropic Gaunt maps (\textit{left}) $g_{\perp}$ and (\textit{right}) $g_{||}$ for source pulsar magnetic field strengths, $B = 10^{11}-10^{13}$ G with discrete sampling for $\omega_c$ and model X-ray photon energies $\omega$ consistent with \cite{MeszarosNagel1985b} for a plasma temperature $kT=8$ keV. In each grid, relatively larger values are represented by brighter shades.}
    \label{color_maps}
\end{figure}

Fig. \ref{color_maps} depicts the values computed in Tables \ref{table:g_perp} and \ref{table:g_para} as color maps for a visual representation of the trend in the figures.

\section{Variation with plasma temperature}

\begin{figure}
    \centering
    \includegraphics[height=.95\textheight]{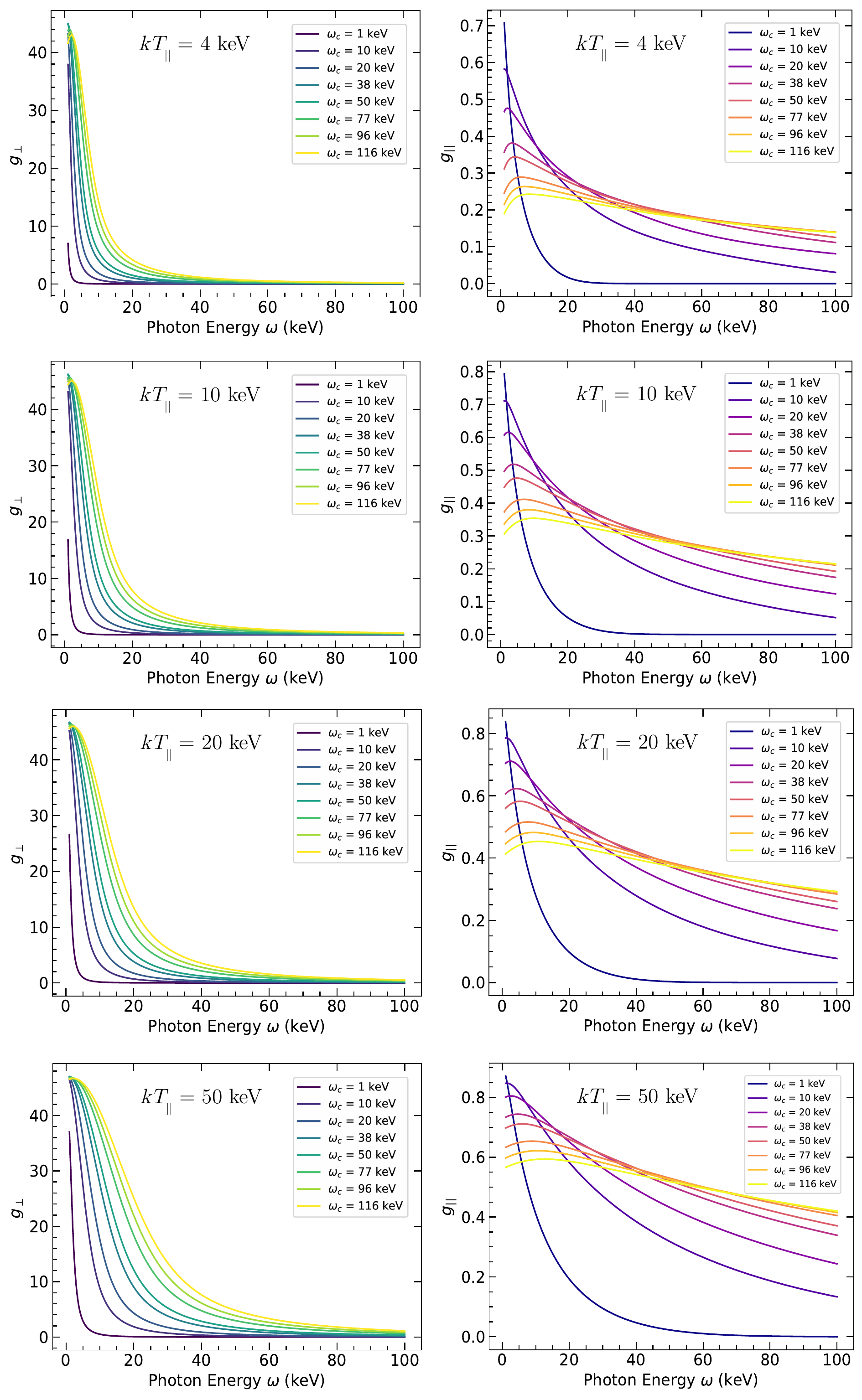}
    \caption{Modified Gaunt curves (\textit{left column}) $g_{\perp}(\omega)$ and (\textit{right column}) $g_{||}(\omega)$ for the full spectral range of X-ray space missions \textit{i.e.} photon energies, $\omega = 1-100$ keV with $\Delta \omega =0.1$ keV with (\textit{isochors}) discrete sampling over the full range of source magnetic field strengths, $B = 10^{11}-10^{13}$ G including model \citeauthor{MeszarosNagel1985b} values for (\textit{top to bottom}) increasing model plasma temperatures of  $kT_{||} = 4, 10, 20$ and 50 keV in \citeauthor{1979A&A....78..136Y}'s range.}
    \label{fig:vary_w_4_10_20_50_keV}
\end{figure}

The variation of the Gaunt factors can also be studied with a change in the third parameter of plasma temperature, $kT_{||}$ \citep{1979rpa..book.....R}. Apart from the model value of $kT_{||} = 8$ keV used by \citeauthor{MeszarosNagel1985b} -- results for which are displayed in Sec. \ref{sec:rnd} -- \cite{1979A&A....78..136Y} suggest electron temperatures of $4-50$ keV for hot strongly magnetized plasma in pulsar atmospheres with an emphasis on the $10-20$ keV range. In an earlier work, \cite{1979ApJ...229L..73Y} discarded temperatures above $10$ keV for Her X-1. Fig. \ref{fig:vary_w_4_10_20_50_keV} shows the change in the anisotropic Gaunt curves with plasma temperatures sampled within \citeauthor{1979A&A....78..136Y}'s range.

The computed results exhibit a variation with the plasma temperature. On comparative inspection, it is observed that the steep fall in the Gaunt curves tends to flatter slopes for hotter plasma. The $g_{||}$ curves in Fig. \ref{fig:vary_w_4_10_20_50_keV} exhibit an increasing upward offset whereas the $g_{\perp}$($\omega_c$ = 1 keV) curve rises to a higher value near the lower X-ray photon energy end ($\omega$ = 1 keV).

\bibliography{sample631.bib}
\bibliographystyle{aasjournal}

\end{document}